\documentclass[conference]{IEEEtran}

\usepackage{amssymb}
\usepackage{xcolor}
\usepackage{subfig}
\usepackage{epstopdf}
\usepackage{graphicx} 
\usepackage[section]{placeins}
\usepackage{url}
\usepackage{longtable}
\usepackage{enumerate}
\usepackage{makecell}
\usepackage{multirow}

\usepackage{tikz}
\usepackage{ctable}

\newcommand\copyrighttext{%
  \footnotesize This work has been submitted to the IEEE for possible publication. Copyright may be transferred without notice, after which this version may no longer be accessible}
\newcommand\copyrightnotice{%
\begin{tikzpicture}[remember picture,overlay]
\node[anchor=south,yshift=10pt] at (current page.south) {\fbox{\parbox{\dimexpr\textwidth-\fboxsep-\fboxrule\relax}{\copyrighttext}}};
\end{tikzpicture}%
}

\begin{document}

\title{Security Apps under the Looking Glass: \\ An Empirical Analysis of Android Security Apps}

\author{
	\IEEEauthorblockN{
		Weixian Yao,\IEEEauthorrefmark{1}
		Yexuan Li,\IEEEauthorrefmark{1}
		Weiye Lin,\IEEEauthorrefmark{1}
		Tianhui Hu,\IEEEauthorrefmark{1}
		Imran Chowdhury,\IEEEauthorrefmark{1}
		Rahat Masood,\IEEEauthorrefmark{1}\IEEEauthorrefmark{2}
		Suranga Seneviratne\IEEEauthorrefmark{1}
% 		Guillaume Jourjon,\IEEEauthorrefmark{3}
% 		Darren Webb,\IEEEauthorrefmark{2} and
% 		Richard Xu\IEEEauthorrefmark{1}
	}\\
	\IEEEauthorblockA{
		\IEEEauthorrefmark{1} University of Sydney, Australia, 
		\IEEEauthorrefmark{2} Data61-CSIRO, \\
		Email: \{first name.last name\}@sydney.edu.au
	}
}

% Weixian Yao, Yexuan Li, Weiye Lin, Tianhui Hu and Imran

% for over three affiliations, or if they all won't fit within the width
% of the page, use this alternative format:
% 
%\author{\IEEEauthorblockN{Michael Shell\IEEEauthorrefmark{1},
%Homer Simpson\IEEEauthorrefmark{2},
%James Kirk\IEEEauthorrefmark{3}, 
%Montgomery Scott\IEEEauthorrefmark{3} and
%Eldon Tyrell\IEEEauthorrefmark{4}}
%\IEEEauthorblockA{\IEEEauthorrefmark{1}School of Electrical and Computer Engineering\\
%Georgia Institute of Technology,
%Atlanta, Georgia 30332--0250\\ Email: see http://www.michaelshell.org/contact.html}
%\IEEEauthorblockA{\IEEEauthorrefmark{2}Twentieth Century Fox, Springfield, USA\\
%Email: homer@thesimpsons.com}
%\IEEEauthorblockA{\IEEEauthorrefmark{3}Starfleet Academy, San Francisco, California 96678-2391\\
%Telephone: (800) 555--1212, Fax: (888) 555--1212}
%\IEEEauthorblockA{\IEEEauthorrefmark{4}Tyrell Inc., 123 Replicant Street, Los Angeles, California 90210--4321}}

% use for special paper notices
%\IEEEspecialpapernotice{(Invited Paper)}

% make the title area
\maketitle
\copyrightnotice
 
\begin{abstract}

Third-party security apps are an integral part of the Android app ecosystem. Many users install them as an extra layer of protection for their devices. There are hundreds of such security apps, both free and paid in Google Play Store and some of them are downloaded millions of times.  By installing security apps, the smartphone users place a significant amount of trust towards the security companies who developed these apps, because a fully functional mobile security app requires access to many smartphone resources such as the storage, text messages and email, browser history, and information about other installed applications. Often these resources contain highly sensitive personal information. As such, it is essential to understand the mobile security apps ecosystem to assess whether is it indeed beneficial to install them. To this end, in this paper, we present the first empirical study of Android security apps. We analyse 100 Android security apps from multiple aspects such as metadata, static analysis, and dynamic analysis and presents insights to their operations and behaviours. Our results show that 20\%  of the security apps we studied potentially resell the data they collect from smartphones to third parties; in some cases, even without the user consent. Also, our experiments show that around 50\% of the security apps fail to identify malware installed on a smartphone.

\end{abstract}

\section{Introduction}
\label{sec:intro}

Smartphones have become an essential element in our daily lives. As the end of 2019, the number of global smartphone users was predicted to reach 3.2 billion, which is a staggering 41\% of the global population~\cite{statista1}\cite{statista2}. Due to the wide range of apps people use in their smartphones, smartphones have become a storage for highly sensitive personal data such as communications, images and videos, health information, and financial details. Also, many enterprises rely on smartphones for their business operations and in some cases even allow their employees to use their own devices for business work. Thus, smartphones have become lucrative targets for cyber attacks and it is of utmost important to keep smartphones secure.

Analogous to antivirus software for computers, mobile security apps are available for smartphones. Many of these mobile security apps are downloaded millions of times and many enterprises usually installs third party security apps or mobile device management software in company-issued phones as an extra layer of security~\cite{verizon}. While indicating that if the users stick to the official app stores and stay up-to-date with the operating system updates, they don't need any mobile security software in their smartphones~\cite{digitasltrends}, both Apple and Google still allow third party security apps in their app stores. Whether the users need to install third-party mobile security apps in their smartphones is a contentious topic.

In multiple occasions some of mobile security apps have been reported for dubious activities such as intentional or unintentional personal information collection, advertisement fraud, or simply providing a false sense of security without actually detecting mobile malware~\cite{wired1}\cite{trendmicro}. The correct operation of a functional third-party-mobile security app requires a significant access to user data and mobile phone resources (i.e. in the from of \textit{permission requests}). For example, a mobile malware detection app will require access to all the stored files to ensure that no malware is transferred to the device~\cite{bitdefender}\cite{avast}. Similarly, a mobile internet security app will require access to all the URLs a user visits in order to prevent the user visiting a malicious website~\cite{fsecure}. As a result, a smartphone user who is thinking of installing a mobile security app needs to make an informed decision whether it is beneficial to install such an app or simply stick to the security measures provided by the platform provider, i.e. majorly Google and Apple. To this end, in this paper we study the functionalities and inner workings of a set Android mobile security apps using metadata analysis, static code analysis, and dynamic behaviour analysis. Specifically, we make the following contributions.

\begin{itemize}
    \item We conduct a keyword search in Google Play Store and identify 328 Android security apps. We conduct an analysis of their metadata to understand the security app ecosystem. We show that despite increasing the security measures in the Android operating system, the number of security apps in the Google Play store is steadily increasing and they are frequently used by the end users. We find that 26 out of 328 apps were downloaded over 10M times and 37 apps were downloaded over 1M times. %We next select 100 apps from  all popularity levels for further investigation.

    \item We examine the privacy policies of 100 selected security apps and found that 55 of them may decide to share the data they collect with legal authorities or business affiliates. We also found that 20\% resell the data to third parties, in some cases without user permission. Nearly, 70\% of apps collect sensitive user information such as installed apps and device information.    %either with the purpose of improving services, for advertisement purposes, or to inform users about security issues. 

    \item Using static analysis we found that on average a security app requests approximately 22 permissions that include approximately four dangerous permissions. We observed that \texttt{WRITE EXTERNAL STORAGE}, \texttt{READ EXTERNAL STORAGE}, and \texttt{READ PHONE STATE} are the most frequently requested dangerous permissions. While these permission requests are essential for the correct functionality of the security app, they can as well be easily abused. At wrong hands, such level of information access together with the privacy consent users give during installation can have serious consequences.
    
    %\textcolor{red}{features --- different from dangerous permissions?}
    
    %All these features are considered critical given their nature of reveal sensitive information about a user e.g. user location trajectories. 
    
    %We perform extensive evaluation on apps security permissions and features as part of static analysis. We found that on average, each app requests 24.75 permissions. The average number of normal permissions is  20.29,  which  was  followed  by  4.18  dangerous  permissions and  0.28  signature  permission. We also observed that ‘WRITE\_EXTERNAL\_STORAGE’, ‘READ\_EXTERNAL\_STORAGE’ , and ‘READ\_PHONE\_STATE’ are the three dangerous permission that have are requested by more than 65 apps. 
    
    \item We measure the effectiveness of the security apps in malware detection by copying and installing known malware samples into a test phone and checking whether security apps can detect them. We found that an alarming number of the apps fail to detect the presence of a malware. For example, only $\sim$20\% of the apps were able to detect malware copies stored in a phone, and only $\sim$50\% of the apps were able to detect malware installed in a phone. Also, we note that another significant fraction of security apps were not having up to date virus databases and as such fail to identify recent malware samples.

    %Our results indicate that the rate of detection for on-device (copied) malware is 20\% less as compared to installed malware. The on-device malware is detected by only 19 apps on average, whereas, the installed malware is detected by 44 apps on average.  The lowest detection rate is shown by the Banking Malware which was recently introduced in 2019. This indicates that antivirus apps do not update their virus detection frequently and is subject to high false positives. On the other hand, we observe that Xbots is the malware category that is detected by antivirus apps without even installation, whereas, Trojans has the highest detection rate after APK installation.
    
    % over 60 antivirus \textcolor{red}{be specific how many apps? over 60 usually doesn't mean anything} apps did not identify on-device malware APK files. Similarly, installed malware were detected only by 50\% of the apps \textcolor{red}{check the number}. All in all, 35\%  of  all  selected  anti-virus apps do  not  provide  any  virus  detection  functions at all providing only a false sense of security. \textcolor{red}{How does this 35\% come from - combination of above two?}
    
    \item Finally, we analyse the network traffic generated by the security apps, and show that while majority of the traffic going out are encrypted, still there is unencrypted communication using port 80. We also analysed the type of information security apps sent to their remote servers and found that ‘device model’, ‘root build ID’, and ‘brand’ are the most frequently sent out information. Moreover, we also found evidence of sensitive personal information in the likes of installed apps and persistent identifiers (e.g. Android ID and BSSID) is being collected. 

% We found one app that contact 180 different IP address distributed over 100 domain. 
% \footnote{\textcolor{blue}{We assume this purpose by analyzing privacy policies of each app.}}.

\end{itemize}

The rest of the paper is organised as follows. In Section~\ref{Sec:Related} we describe related work and in Section~\ref{Sec:Dataset} we describe our data collection methodology and present a basic characterization of the security apps. Privacy policy analysis and permission analysis are presented in Section~\ref{Sec:PrivacyPolicy} and Section~\ref{Sec:Permissions} respectively while Section~\ref{Sec:Tracking} delves into the network traffic generated by security apps. We discuss the implications of our findings, limitations, and concluding remarks in Section~\ref{Sec:Discussion}.

\section{Related Work}
\label{Sec:Related}

There is a plethora of work studying characteristics and behaviours of mobile apps. We present related work under three topics; { \textit{i) mobile malware, over-permissions and other dubious behaviours detection, ii) behaviour characterization of one specific group of apps}}, and { \textit{iii) empirical studies of antivirus apps}}.

\subsection{Mobile malware, over-permissions, \& dubious behaviors}
When it comes to mobiles apps, malware detection is a topic that has been studied extensively~\cite{zhou2012dissecting,grace2012riskranker,wu2012droidmat,sahs2012machine,mclaughlin2017deep,suarez2017droidsieve}. Early work by Zhou and Jian~\cite{zhou2012dissecting} was the first to study Android malware samples in the wild. Authors characterised over 1,000 malware samples belonging to 49 malware families, evolved during the early years of Android. Also, authors tested the detection rates of four mobile malware detection software and showed that in best cases 79.6\% samples were detected and in the worst case only 20.2\% samples were detected. Multiple subsequent work proposed methods to detect mobile malware. Such methods include static analysis~\cite{zhou2012hey,hoffmann2013slicing}, dynamic analysis~\cite{yan2012droidscope,enck2014taintdroid}, and hybrid methods~\cite{rasthofer2016harvesting,spreitzenbarth2013mobile}. Tam et al.~\cite{tam2017evolution} and Faruki et al.~\cite{faruki2014android} present comprehensive surveys of Android malware analysis techniques.

%X ,,,, .... Initial work highlighted various malware categories in Android environment and subsequent work came up with . proposed methods include static analysis, dynamic analysis, and X . 

In contrast to mobile malware, \textit{apps requesting over-permissions} do not pose a direct threat to the data stored in the phone or the device itself. Rather, these apps have more access to phone resources and data than what is required for the functionality of the app. Such additional access is usually used for advertising and analytics purposes and can be a threat to user privacy. Such access also allows possible future malicious activities. Grace et al.~\cite{grace2012unsafe} and Leandros et al.~\cite{leontiadis2012don} characterised the over-permission behaviours of top Android apps. Gorla et al.~\cite{gorla2014checking} clustered apps based on app descriptions and identified typical API usages for various app functionalities. Then the authors used anomaly detection to identify apps requesting unnecessary permissions. Other comparable work includes~\cite{peng2012using,zhang2013vetting,pandita2013whyper}.

Another body of work focused on detecting various other malpractices in mobile app markets such as spamming~\cite{seneviratne2015early,seneviratne2017spam}, counterfeiting~\cite{rajasegaran2019multi},  and ranking fraud~\cite{surian2017app,zhu2014discovery}. These apps adversely affect the overall health of the Android app ecosystem and in some cases were also reported to contain malware. Majority of the detection methods proposed use various features generated from app metadata like app description, developer information, ratings, and app icons in combination with machine learning models. \textit{In contrast to these works, our work specifically focus on the mobile security apps and study their operations and behaviours.}

\subsection{Behaviour characterization of specific app groups}

Several studies focused on specific functional groups of apps, especially from the privacy and security viewpoint. Meyer~\cite{meyer2019advertising} et al. characterised the advertising behaviours of 135 apps targeting children under the age of five and showed that manipulative and deceptive behaviours are commonplace. Reyes et al.~\cite{reyes2018won} conducted automated analysis of 5,855 most popular free children's apps and found that the majority of the apps are in fact in violation of the Children’s Online Privacy Protection Act (COPPA). For example, 19\% of the apps were found collecting personally identifiable information (PII) and 66\% of the apps were still collecting the persistent device identifiers than the re-settable advertising identifiers. Similar studies has been conducted for health apps~\cite{martinez2015privacy,dehling2015exploring}. More recently, He et al.~\cite{he2020beyond} studied the behaviours of various dubious apps that emerged during the COVID-19 pandemic containing related keywords.   

Ikram et al.~\cite{ikram2016analysis}\cite{ikram2017first} studied the privacy and security aspects of VPN apps and ad-blocking apps that are heavily used by privacy conscious smartphone users and the results were eye-opening. For instance, authors found that 18\% of the VPN apps actually do not encrypt the traffic providing a false sense of confidentiality, and 4\% of the apps contained malware. Similarly, Seneviratne et al.~\cite{seneviratne2015measurement} and Hu et al.~\cite{hu2019want} showed that paid apps can also pose serious privacy and security threats to the users, despite the common belief that such problems predominantly occur in free apps. \textit{Our work is complementary to these studies. We provide an empirical analysis of a security apps; an app group that hasn't received wider attention despite its high importance.}

\subsection{Empirical studies of antivirus apps}

Limited work studied of antivirus software in the context of desktop computers and laptops~\cite{stone2013underground,sukwong2010commercial,rajab2010nocebo}. Rajab et al.~\cite{rajab2010nocebo} measured 15\% of the all malware detected in the web were advertised as free antivirus software by analysing over 200 million malicious web pages. Stone-Gross et al.~\cite{stone2013underground} delved into the economics of the scare-ware and fake antivirus market and showed that the organized underground business of fake antivirus software easily reaches the scale of hundreds of millions of dollars. Finally, Sukwong et al.~\cite{sukwong2010commercial} evaluated the effectiveness of the commercial antivirus software and found that they could not identify all the modern threats. 

%Antivirus apps has been under scrutiny in other operating systems. In as early as 2000s Windows antivirus software ... 

\textit{Our work is comparable to the industrial report~\cite{fedler2013effectiveness} and the studies by Rastogi et al.~\cite{rastogi2013droidchameleon} and Zheng et al.~\cite{zheng2012adam}. However,  these studies solely focus on the effectiveness of the Android security software, whereas our study is multi-faceted. We not only study the effectiveness of the security software, but also investigate various other privacy and security threats they can pose.}

\section{Data set and Basic Characteristics}
\label{Sec:Dataset}

In this section, we describe our data collection and pre-processing steps. We also, provide a basic characterisation of security app metadata.

\subsection{Data collection}
We first conducted a keyword search in Google Play store with security related terms such as ``antivirus'', ``malware'', and ``security'' and identified 328 potential Android security apps. Next, we crawled the Play Store pages of these apps and downloaded the executable file (APK). We also recorded the metadata such as the app description, number of downloads, ratings, and developer information. 

%We then sorted the downloaded apps according number of downloads, rating count, and average star rating as proposed in~\cite{seneviratne2015early}; an app raking heuristic to identify top apps.

%The first order lists apps with highest number of downloads whereas the second and third order lists apps according to rating count and average star, respectively~\cite{benck2014taintdroid}. 

Since some of the analysis we subsequently conduct requires manual effort, we next select a representative sample of 100 apps out of the 328 apps we downloaded. First, we ranked the apps based on the number of downloads, number of user reviews, and rating, similar to what was done in previous work~\cite{seneviratne2015early,seneviratne2017spam}. Next, we divided 328 apps into three groups of 100, 100, and 128 apps, respectively. From the first group, we selected top-40 apps, and that included well known security apps such as McAfee and Kaspersky. We randomly selected 30 apps each from the other two groups. This selection gives us a sample of 100 apps that contains highly popular security apps as well as apps that are not so well known.

%Next, we select 100 apps out of \textcolor{red}{328} for further analysis. We choose the number 100 because \textcolor{red}{...}. For this purpose, we divide 328 apps into three blocks of 100, 100, and 128 apps, respectively. From the first block of 100, we select the top 40 apps which include apps from antivirus companies such as McAfee and Kaspersky. We randomly selected 30 apps each from the other two groups. This selection gives us a sample of 100 apps that contains highly popular security apps as well as apps that are not so well known from the .....

%30 apps from each of the other two blocks to remove biasness from the data. This is because the most popular apps may be well-designed and thus unable to reflect the cases of other antivirus apps.

\subsection{Metadata Analysis}

\begin{figure}[!htbp]
\centering
%\label{fig:phished_web}
      \includegraphics[scale=0.45]{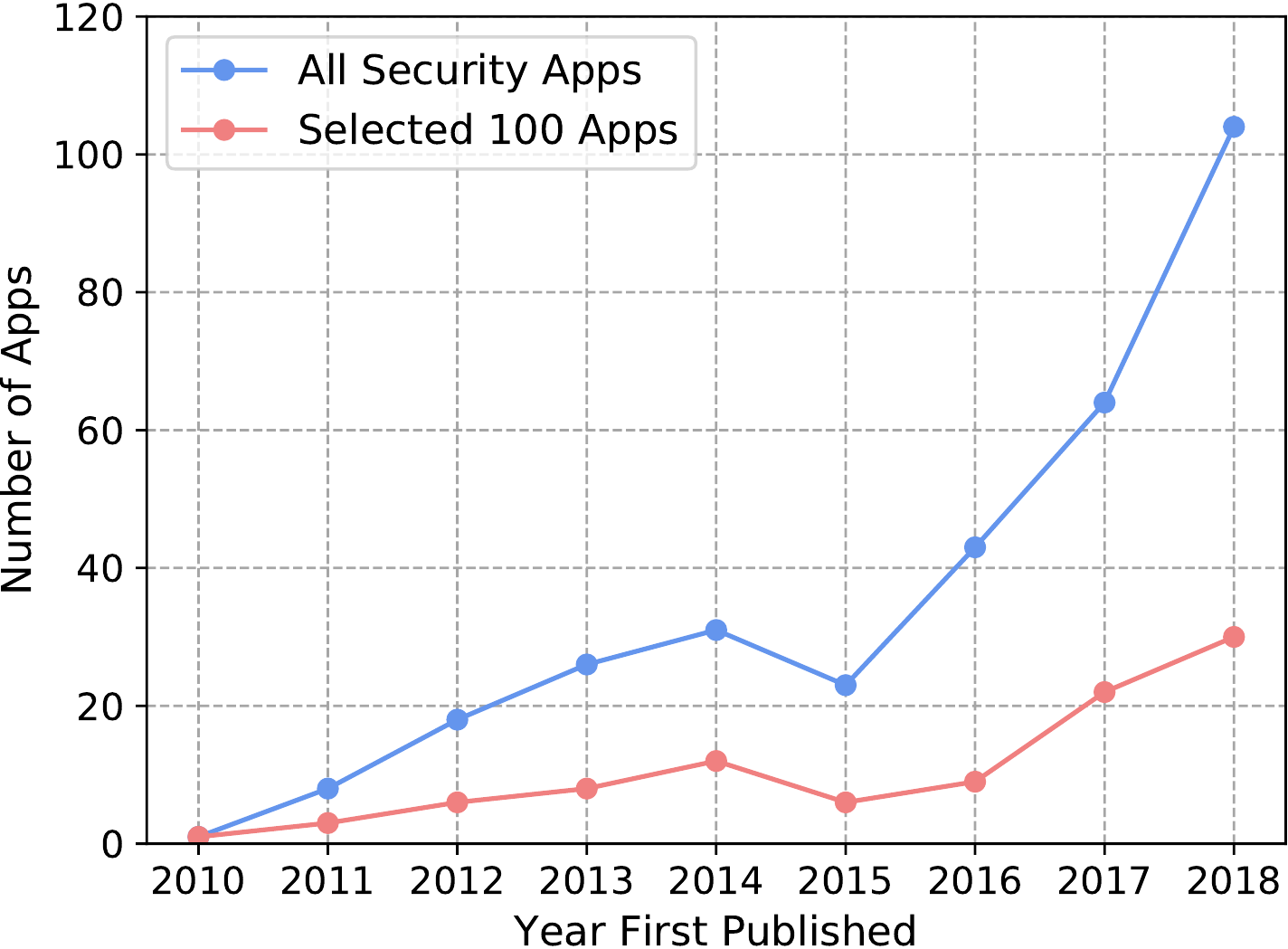}
\caption{Timeline of security apps in Google Play Store}
\label{fig:fig1}
\end{figure}

\noindent{\bf Timeline of Security Apps:} To identify the trends in security app availability, we extracted the \texttt{creation time} from the metadata of each app. Figure~\ref{fig:fig1} shows the development timeline of 328 antivirus apps as well as 100 apps that we selected. We notice a steady increase in the number of antivirus since 2010, and a steep increase since 2015. Overall, it indicates that there is an increasing demand for security apps attracting more developers. \\ \vspace{-2mm}

% which indicates an increasing focus on the importance of antivirus apps. We see similar trend in across 100 selected apps which proves the preciseness of our way to select data. \\ \vspace{-2mm}

% \begin{figure}[!t]
% \centering
% \label{fig:phished_web}
%       \includegraphics[width=\columnwidth, keepaspectratio]{figures/Both_Creation_Time.pdf}
% \caption{Timeline of Antivirus Apps at Google Play Store \textcolor{red}{TODO-Need to Add possible explanation of why there are 319 apps instead of 328 apps?}}
% \label{fig:1}
% \end{figure}

\begin{figure}[!htbp]
\centering
%\label{fig:phished_web}
      \includegraphics[scale=0.45]{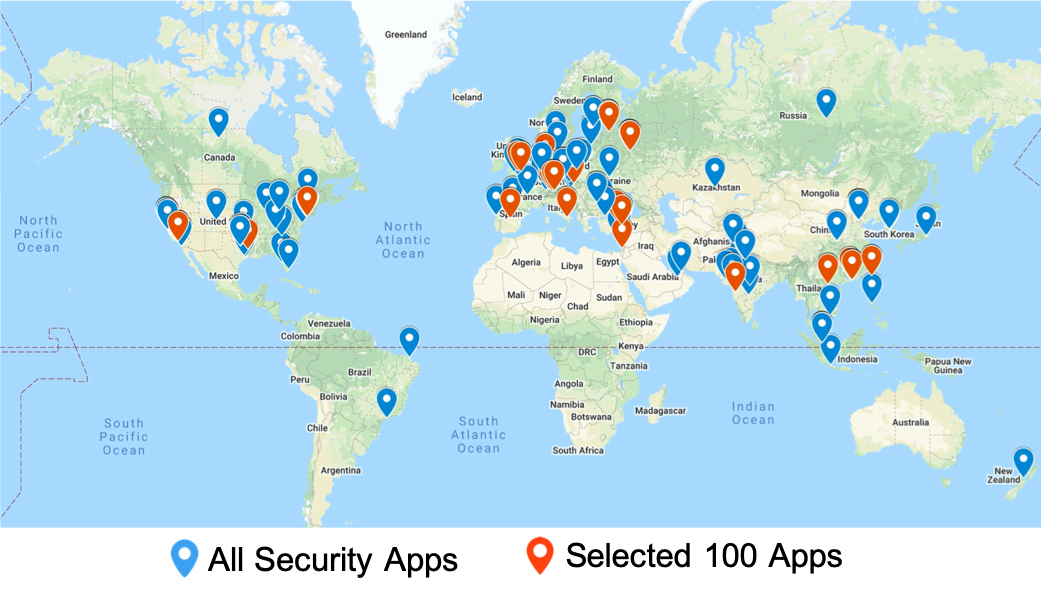}
\caption{Geo-locations of the security apps}
\label{fig:2}
\end{figure}

\noindent{\bf Geographical Location:} We next analyse the geographical location of antivirus apps as illustrated in Figure~\ref{fig:2}. According to the figure, most of these apps are operating from the United States, Europe, China, and India.  However there are some notable outliers. For example, well known Kaspersky security app is the one marked in Russia. One listed under New Zealand is ``Emsisoft Mobile Security'' - a security app that provides services such as malware scanning, anti-theft, privacy advisor, and web security. Similarly, apps in Indonesia and Philippines include 	
Hackuna - Anti-Hack (a WiFi security app) and Search Phone Security - Booster \& Cleaner,\footnote{This app is no longer available in Google Play Store}  respectively. \\ \vspace{-2mm}

% The Red marker indicates the Selected 100 Apps whereas the Blue markers indicates all selected apps, respectively. \textcolor{red}{TODO - Figure is good. Can we add the legend at the bottom of the figure. You can do this manually in Powerpoint. This will save caption being awkwardly long }

\begin{figure}[!htbp]
\centering
%\label{fig:phished_web}
      \includegraphics[scale=0.45]{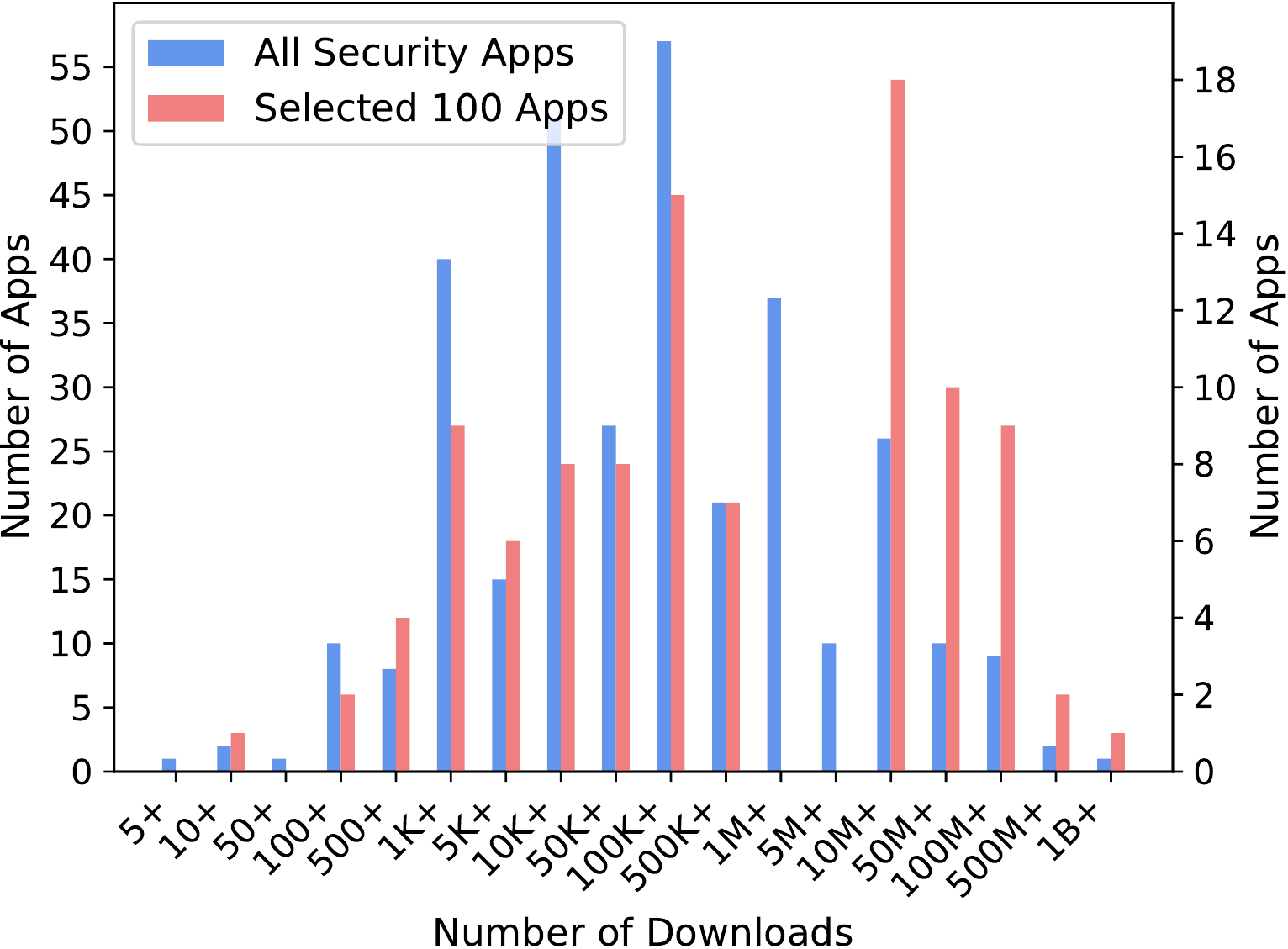}
\caption{Download distribution of the security apps}
\label{fig3}
\end{figure}

\noindent{\bf Popularity of Antivirus Apps:} Figure~\ref{fig3} shows the download distribution the security apps. One app ``Clean Master'' was downloaded over one billion times. It is a multi-purpose security app providing variety of services  such as app lock, battery saver, junk file cleaning, virus detection, and Android optimization. Nonetheless, this app was recently removed from Google Play Store under the suspicion that it is involved in an advertisement fraud scheme~\cite{mobilemarketer}. This indeed is a compelling incident to highlight how easily security apps can abuse the trust placed by their users.

``Device Care'' by Samsung that provides device maintenance services for Samsung mobile devices was downloaded over 500 million times. The app analyses battery usage on a per-app basis and identifies battery draining apps. It also offers features such as power saving mode, remove unnecessary files, RAM management, malware detection, and device configuration. Potentially this app might be pre-installed in some Samsung devices, which can be a reason behind the higher number of downloads. Another app that had over 500 million downloads is ``CM (Cleanmaster)''. This app was also from the same developer as the previous ``Clean Master'' app and was recently banned from Google Play Store.  Finally, we note that there are no selected apps in the download range from one million to ten million because after the ranking we did (cf. Section III-A), from the first 100 apps we selected only the top-40 apps whereas from the other two groups we selected apps randomly.   \\ \vspace{-2mm}  
% \textcolor{red}{TODO Other apps that had over 500 million downloads include X....}. 
% \textcolor{red}{TODO - Rahat: I think it is better to merge these two graph. Way to do is you insert another axis from the right 0-20 and in a different color plot the columns of the 100 apps. Also talk about the apps having higher downloads, what was downloaded 1B times what were 500M times? } 

% 1\begin{figure}[!t]
% \captionsetup{skip=0pt, justification=centering}
% \centering
% \subfloat[All antivirus apps] {\label{fig:328_downloads}
%       \includegraphics[width=\columnwidth, keepaspectratio]{figures/328_Downloads.pdf}} \\
% \subfloat[Selected 100 antivirus apps] {\label{fig:100_downloads}
%       \includegraphics[width=\columnwidth, keepaspectratio]{figures/100_Downloads.pdf}}
% \caption{Download Distribution of Antivirus Apps at Google Play Store}
% \label{fig:3}
% \end{figure}

\begin{figure}[!htbp]
\centering
%\label{fig:phished_web}
      \includegraphics[scale=0.45]{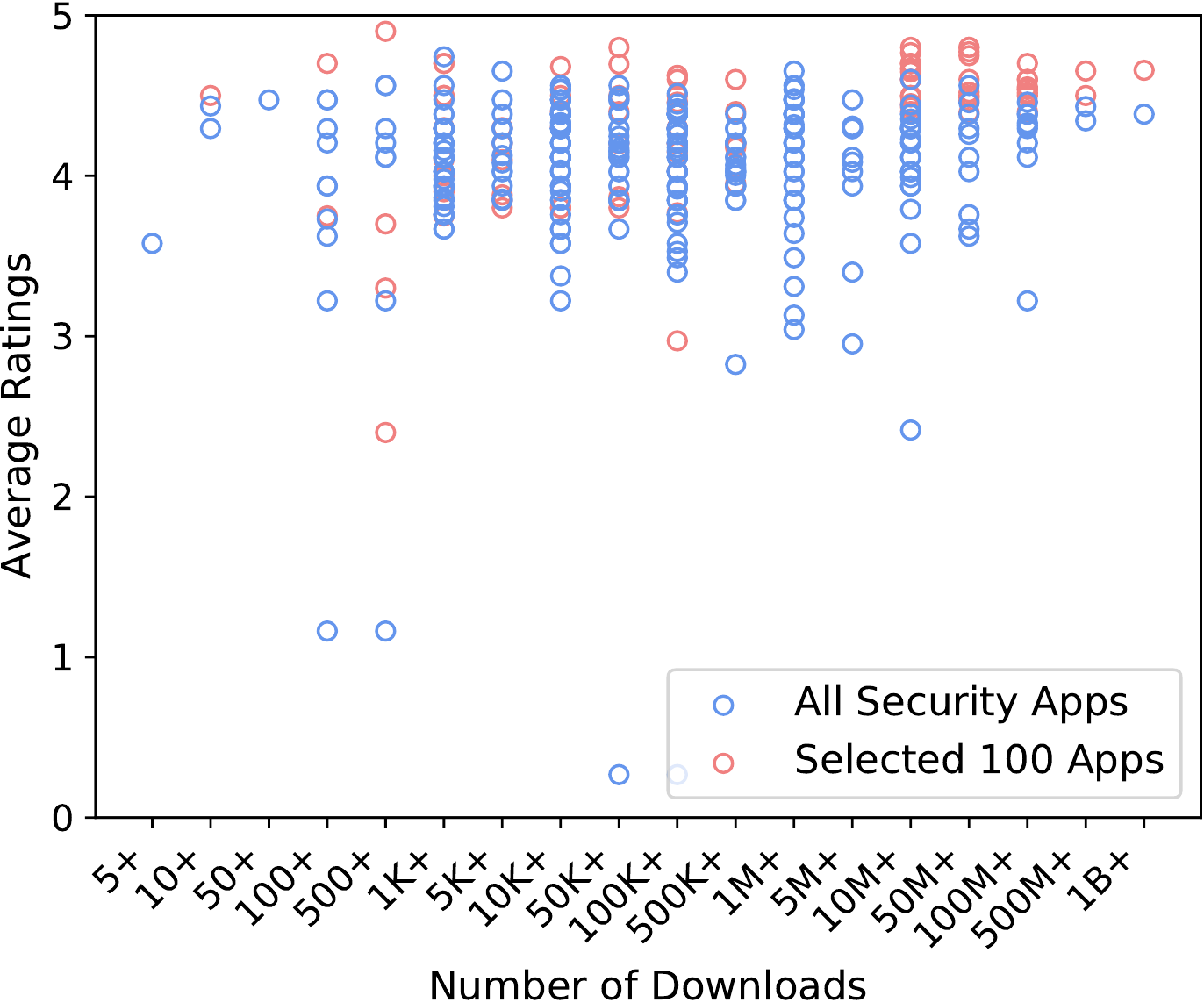}
\caption{Rating distribution against the number of downloads}
\label{fig:4}
\end{figure}

\noindent{\bf Number of Downloads vs. Ratings:} We next analyse the distribution of ratings (ranging from 0 to 5) against the number of downloads. As shown in Figure~\ref{fig:4}, the vast majority of the security apps have a very high rating (over 4), indicating they are well received by the end users. For example, all the apps with over 10 million downloads had a rating higher than 4.4. The banned ``Clean Master'' app had an average rating of 4.6. %\textcolor{blue}{A notable anomaly is the app ``Kaspersky Antivirus AppLock \& Web Security'' which had over 50k downloads but a very low average rating.} \textcolor{red}{TODO I don't think this is right. I one in the figure has an avergae rating close to zero. This one has 4.8 in Google Play now.} \\ \vspace{-2mm}

%, whereas apps with lower number of downloads have dispersed ratings. This trend is similar for the selected 100 apps as well as all 328 apps, as shown in Figure~\ref{fig:4}. For instance, \texttt{com.cleanmaster.mguard} which is downloaded 1B+ times has an average rating 0f 4.6 whereas the apps with 100+ downloads have average rating ranging from 4.5 to 1.2, respectively.
% \textcolor{red}{TODO-Rahat: I think in this format these two figures do not give useful information. May be rating against number of reviews makes more sense. We could arugue that higher number of reviews and high average rating means the app is actually popular. Also, again one graph for bot in two colors.}  

% \begin{figure}[!t]
% \captionsetup{skip=0pt, justification=centering}
% \centering
% \subfloat[All antivirus apps] {\label{fig:328_downvsrate}
%       \includegraphics[width=\columnwidth, keepaspectratio]{figures/328_DownVsRate.pdf}} \\
% \subfloat[Selected 100 antivirus apps] {\label{fig:100_downvsrate}
%       \includegraphics[width=\columnwidth, keepaspectratio]{figures/100_DownVsRate.pdf}}
% \caption{Rating Distribution with respect to Number of Downloads}
% \label{fig:4}
% \end{figure}

\subsection{App description mining}

Though all the apps we selected broadly fit into the category of Android security, they have different security functionalities such as ``internet security'', ``malware scanning'', or ``virus removal''. To understand what types of services and functionalities the security apps provide, we next mine the app descriptions of the 100 apps we selected. Our approach is to use the \textit{doc2vec} model~\cite{le2014distributed} to obtain vector representations of app descriptions and cluster them using \textit{k-means} clustering to identify functional groups. \\ \vspace{-3mm}

\noindent{\bf doc2vec model} - Since a large text corpus is required to train the doc2vec model, we start with 25,000 random app descriptions we downloaded from Google Play Store as a part of our previous work~\cite{rajasegaran2019multi}. We pre-process the app descriptions with standard natural language processing techniques such as removing symbols, non-English characters, stop word removal, and stemming. Next, we train a doc2vec model using the resulting corpus that produces a vector representation of size 100 for a given text.  \\ \vspace{-3mm}

\begin{figure}[!htbp] 
  \subfloat[Cluster 1]{% 
    \includegraphics[scale=0.25]{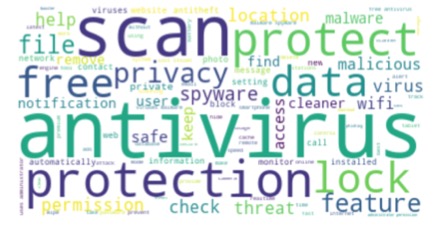} 
    \label{fig:01}
  } 
  \subfloat[Cluster 2]{% 
    \includegraphics[scale=0.25]{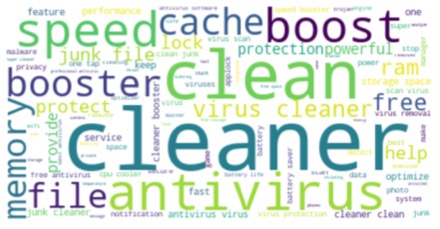} 
    \label{fig:02}
  } \vspace{-2mm}
  \caption{Main themes of Android security apps}
  \label{fig:0}
\end{figure}

\noindent{\bf k-means clustering} - We used k-means clustering to cluster the 100 selected apps based on the vector representations produced by doc2vec. We identified the optimal number of clusters based on the \textit{silhouette score}, which is a standard metric used to measure the quality of clustering. The range of silhouette coefficient is [-1,1]. A higher value of silhouette score implies that there is high similarity between the elements of individual clusters and the clusters are well separated. 

We calculate silhouette score for different $k$ values ranging from 2 to 10.  The highest value of silhouette score resulted when $k$ was two. According to this clustering, Cluster-1 contained 65 apps whose major focus is on providing virus scanning functions and included keywords such as ``scan'', and ``protection''. Cluster-2 had 35 apps having keywords in the likes of ``cleaner'' and ``boost''. These apps are mainly tools that facilitate removing unwanted files and cleaning up memory with extra features for security and virus detection. We highlight the main keywords in the two clusters in Figure~\ref{fig:0}.

\section{Privacy Policy Analysis}
\label{Sec:PrivacyPolicy}

\begin{table*}[t!]
\centering
\caption{Analysis summary of privacy policies } \vspace{-4mm}
\begin{tabular}{p{3.3cm}|p{1.5cm}|p{2.6cm}|p{2.3cm}|p{2.3cm}|p{2.3cm}}\specialrule{.12em}{1em}{0em}

 \multicolumn{2}{c|}{{\bf Type of data}} &   \multicolumn{2}{c|}{{\bf Use of data}} &  \multicolumn{2}{c}{{\bf 
 Data shared with third-parties}} \\ \hline

{\bf Attribute } & {\bf No. of Apps} & {\bf Attribute } & {\bf No. of Apps} & {\bf Attribute } & {\bf No. of Apps} \\ \hline

Personal information & 71 & Newsletter & 53 & Legal authorities & 55 \\ 
User behavior & 62 & Service improvement & 77 & Business affiliates & 48 \\ 
Hardware information & 75 & Security services & 48 & Business partners & 55 \\ 
Upload files to cloud & 30 &  &  & Re-sellers & 21 \\ 
 &  &  &  & Others & 30 \\ 

\specialrule{.12em}{0em}{0em}
\end{tabular} \vspace{-2mm}
\label{tab:1}
\end{table*}

As mentioned earlier in the introduction, smartphone users place an enormous amount of trust on the developers of security apps by entrusting them with their most personal data. Thus, it is important to understand whether there are any additional uses of data and resources accessible to security apps apart from the sole purpose of keeping user devices safe. Therefore, we next investigate the contents and the terms of the privacy policies of the 100 selected security apps. Surprisingly, five apps didn't have any privacy policy. The most notable was ``See Who Is Tracking You'', an app that allows the users to keep track of other apps collecting their location information in background. Currently over 500,000 users have installed this app without realising that there is no privacy policy.

%\textcolor{red}{TODO - What are those apps? Are they downloaded heavily? Can you list the app id and number of downloads. If it is interesting I will add some text here. Also, here it say 94 and below it says 95 - what is the correct number. For example, if there was an app downloaded millions of times with out a privacy policy we should report that. Later you mention only a single app did not have a privacy policy.}
% \textcolor{blue}{ The correct number is 95}.

% \textcolor{blue}{com.omelettersstudios.TrackingAlert: 500,000+(2020)
% + The privacy document is not listed on the Google Play store, we are not sure if it has the document}.

% \textcolor{blue}{com.androhelm.antivirus.free: 500,000+(2020); This app does have a the privacy policy, but it is too generally, does not clarify clearly what information they will collect nor how they use the data.}.

% \textcolor{blue}{com.max.maxantivirus: 100+(2020); When we worked on the privacy work, the privacy policy link listed on the Google Play store is invalid, but now the link is working, maybe they update the content for privacy.}.

% \textcolor{blue}{com.appsplan1.freespywareremovalinfo: 10,000+(2020); Do have a privacy policy link, but page no found.}.

% \textcolor{blue}{org.stel.sProtect: 1,000+(2020); this app does not collect any data because it claims that they do not take any permission to read/write sensitive data such as contact information record}.

We visited the ULRs of the privacy policies defined in the home pages of the apps and collected a copy of their privacy policies. Next, we read the privacy policies ourselves searching answers for three high level questions: 

\begin{enumerate}[i)]
\item {\bf \textit{What type of data the security apps collect?}} Since security apps have access to highly sensitive private data, it is important to understand whether they actually collect such data. Users have the right to know what data is collected and usually this information is hidden in the fine-print of the privacy policies. We could categorise the such data collections into four high level categories; \textit{personal information} (e.g. personally identifiable information such as name, email, and persistent identifiers), \textit{user behaviour} (e.g. how the app is used or other apps' usage data), \textit{hardware information} (e.g. device model and brand), and \textit{file upload} (e.g. obtaining copies of stored data).

\item {\bf \textit{What are the intended uses of collected data?}} - Usually apps collect data for
legitimate purposes such as measuring user experience, identifying bugs, or identifying feature requirements. While such data collections are understandable, many privacy regulations stipulate that app developers must clearly explain how the collected data is used. Thus, it is important to understand whether security apps comply to such legal requirements. We categorised the uses of data into three; \textit{newsletter} (e.g. registering users into the developers mailing lists to receive information and promotional material), \textit{service improvements} (e.g. identifying bugs), and \textit{security services} (e.g. the core services provided by the app such as virus scanning)

%the customer experience improvement and app's function update purposes. App company should let users know how do they use users' information. However the privacy policies are expected to explicitly include how the collected data is used. 

\item {\bf \textit{Is the data being shared with third parties? If yes, who are the third parties with such access}} - In some occasions mobile app developers may  share the data they collect with third parties. It is necessary for security app users to understand whether this is happening to their data and under what circumstances. It is equally important to understand who the above third parties are. For example, data can be shared \textit{offline} in a business-to-business manner with analytics companies. Also, data can be shared \textit{online} through embedded third-party advertisement and analytics libraries. Finally, data can be also shared \textit{on-demand} with bodies such as law enforcement. We categorise such third parties into four; \textit{legal authorities} (e.g. law enforcement requesting user data), \textit{business affiliates}, \textit{business partners}, and \textit{re-sellers} (various business partners of the app developer). We also find occasions where the associated third party information is unclear. Such cases were categorised into \textit{other}.

%offline  or online (e.g. via advertising or analytics SDKs) with third parties. Also, they may have provisions share such data on an on-demand bases with third party bodies such as law enforcement. Thus, 

% Usually, the date collected from the users should be stored securely in the company and should not access without the permission. Nonetheless, it has been seen in multiple occasions that collected data is shared with third parties for purposes such as advertising, user tracking, or in some cases with law enforcement. 

%\item {\bf \textit{Who are third parties with access to the collected data?}} - 

%Some apps claim that the data they collected will be shared with their sub-companies or legal authorities. Users deserve to know where are the date going

\end{enumerate}

We adhered to manual reading than automated analysis because usually the privacy policies are complex legal documents with low readability scores~\cite{mcdonald2009comparative} and as of now there is no automated tool that can be readily used. We tested the tool Polisis~\cite{harkous2018polisis},\footnote{https://pribot.org/polisis} nonetheless it was not able to conclusively provide reliable answers to the questions we were after.

In Table~\ref{tab:1} we summarise our findings and the detailed analysis of each app policy with references to the paragraphs and the sections of privacy policies is available online.\footnote{\url{https://tinyurl.com/ya8rhxgz}} We recommend interested readers to go through the table provided in the link for further information on app policies.  We found that 55 apps may share the data with legal authorities if required and almost all the apps had some form of data sharing with their business partners. Around 70 apps were found to be collecting personal information and hardware information.

\section{Permission Analysis}
\label{Sec:Permissions}

%\subsection{Apps Security Analysis}

We next analyse the permission usage of the selected apps. In Android operating system apps can only access phone resources be it hardware or data, only after declaring their intentions as permission requests in the Android manifest XML file associated with the app. According to Android documentation~\cite{AndroidPermissions} there are three permission categories; normal permissions, dangerous permissions, and signature permissions. The \textit{normal permissions} though they have to be declared by the app developer in the Android manifest file, do not require user's explicit consent before accessing the data or the resource. Examples for normal permissions include access to the Internet, WiFi state, and detecting phone reboots. In contrast, \textit{dangerous permissions}  require user's explicit consent before accessing the data and hence pop up a dialog window to request access when the app needs that permission for the very first time. Examples for dangerous permissions include access to contacts, storage, or location. Finally, \textit{signature permissions} are a special type of permissions that is granted at installation time only if the app that attempts to use a permission is signed by the same certificate as the app that defines the permission. Signature permissions are usually used when a single developer developing a suite of apps.

%\textcolor{red}{for instance it include permissions such as ......} On contrary, critical permissions require user's consent before accessing the data and hence pop up a dialog window to request access. \textcolor{red}{These type of permissions include ....}  The signature permissions are a special type that require permission from a mobile system to access resources. \textcolor{red}{These include...}

\begin{figure}[!htbp]
\centering
%\label{fig:phished_web}
      \includegraphics[scale=0.45]{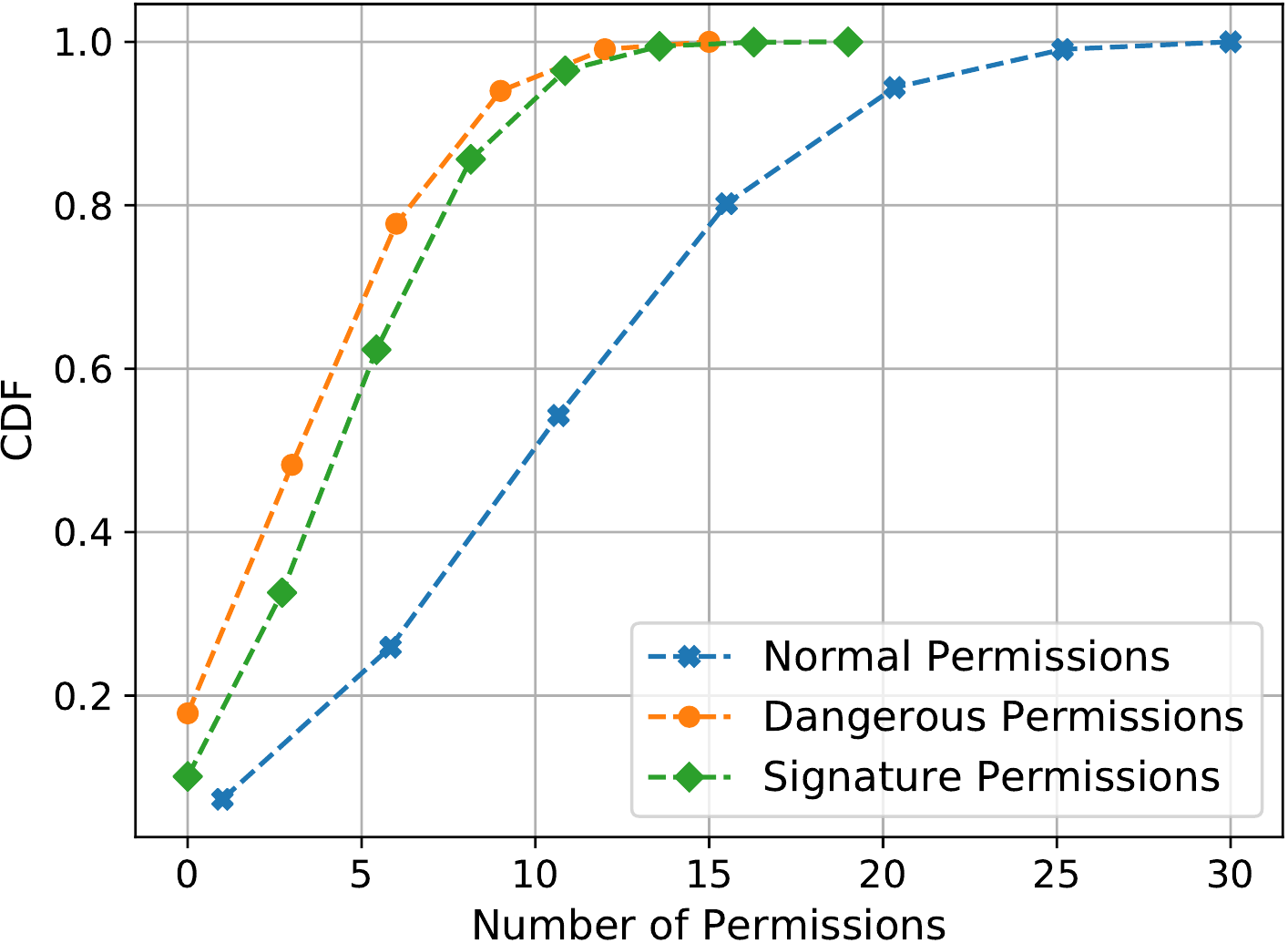}
\caption{Cumulative distribution function of number of permissions}
\label{fig:CDF_permissions}
\end{figure}

The permissions requested by an app usually can give an understanding of the features of the app as well as any associated privacy and security risks. In order to check the permission requests, we decompiled each Android APK executable and extracted the \texttt{AndroidManifest.XML}. By parsing the Manifest file we extracted the permission requests defined by each app. We observed that, on average a security app requested 22.09 permissions. Out of that average number of normal permissions was 12.23, followed by 5.52 signature permissions and 4.33 dangerous permissions. `McAfee Security'' requested the highest number of permissions by an app, which was 51. In Figure~\ref{fig:CDF_permissions}, we show the cumulative distribution of different apps. We notice that approximately 60\% of the apps requested five dangerous permissions.

% \textcolor{red}{Rahat can you put the correct app and the number of permissions}. 

%the ``Manifest'' file of each APK to get the information about permissions and features requested by the apps. Hence, the below analysis is based on the decompiled data of 100 antivirus apps.

% \subsubsection{Permissions Analysis}
% Apps require permissions to access mobile users' data or phone settings. there are three categories of permissions: normal permissions, signature permissions, and critical permissions.  Figure~\ref{fig:5} shows the distribution of permissions in top 10 apps and all the apps. Each permission type is represented in different colors.     

% The range of the number of permissions is between 1 and 100. In contrast to the number of special permission and signature permission, majority of them request a large number of normal permissions. On average, each application request 24.75 permissions. The average number of normal permissions is 20.29, which was followed by 4.18 dangerous permission and 0.28 signature permission.  The number of permission of an application called “hyperspeed.rocketclean” is 100, the largest one,  while the reverse is true for an application called “org.stel.sProtect” , with the figure being one only, the smallest one.

%  \textcolor{red}{TODO - Rahat, better to make x axis as number of permissions and show the cumulative number of apps. Also use dangerous permissions than critical permissions}

% \begin{figure}[!t]
% \captionsetup{skip=0pt, justification=centering}
% \centering
% \subfloat[Permission Distribution of Top 10 Apps] {\label{fig:top_10_permissions}
%       \includegraphics[width=\columnwidth, keepaspectratio]{figures/Bar_Top10_Permissions.pdf}} \\
% \subfloat[CDF of Permissions in All Apps] {\label{fig:CDF_permissions}
%       \includegraphics[width=\columnwidth, keepaspectratio]{figures/CDF_Permissions.pdf}}
% \caption{Permissions Distribution of Apps}
% \label{fig:5}
% \end{figure}

\subsection{Normal Permissions} 

\begin{figure}[!htbp]
\centering
\includegraphics[scale=0.45]{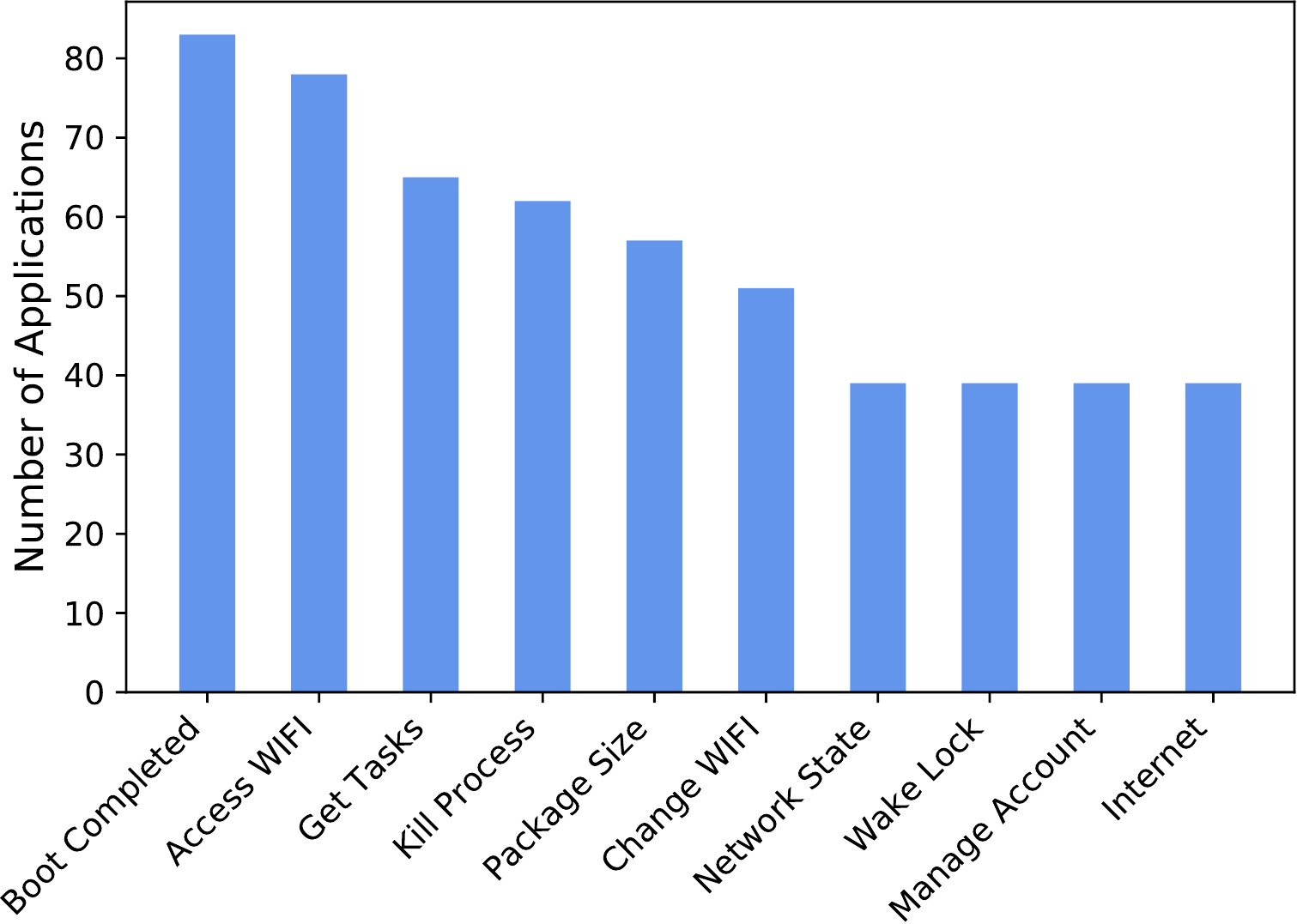}
\caption{Top 10 Normal Permissions.}
\label{fig:Normal_top_10_permissions}
\end{figure}

We observe 149 different normal permissions were requested by the security apps. In Figure~\ref{fig:Normal_top_10_permissions} we show the 10 most frequently requested permissions. \texttt{RECEIVE\_BOOT\_COMPLETED} is the most frequently sought permission which is understandable given that security apps needs to continuously run in the background across device reboots. Other frequently requested normal permissions are predominantly related to the internet access. 

%, it can be clearly seen that the ``READ\_SETTING'' permission is requested 283 times, followed by 213 times of ``WRITE\_SETTING'' permission. The system would allow applications with The ``READ\_SETTING'' permission allows apps to read mobile settings whereas,the ``WRITE\_SETTING'' permission enables apps to change mobile system settings, for instance, control the volume.

% \begin{figure}[!t]
% \captionsetup{skip=0pt, justification=centering}
% \centering
% \subfloat[Distribution of Top 10 Normal Permissions] {\label{fig:Normal_top_10_permissions}
%       \includegraphics[width=\columnwidth, keepaspectratio]{figures/Bar_Top10_Normal_App_Permissions.pdf}} \\
% \subfloat[Distribution of All Normal Permissions] {\label{fig:Normal_CDF_permissions}
%       \includegraphics[width=\columnwidth, keepaspectratio]{figures/Bar_All_Normal_App_Permissions.pdf}}
% \caption{Distribution of Normal Permissions}
% \label{fig:6}
% \end{figure}

\subsection{Dangerous Permissions} 

\begin{figure}[!htbp]
\centering
%\label{fig:phished_web}
      \includegraphics[scale=0.45]{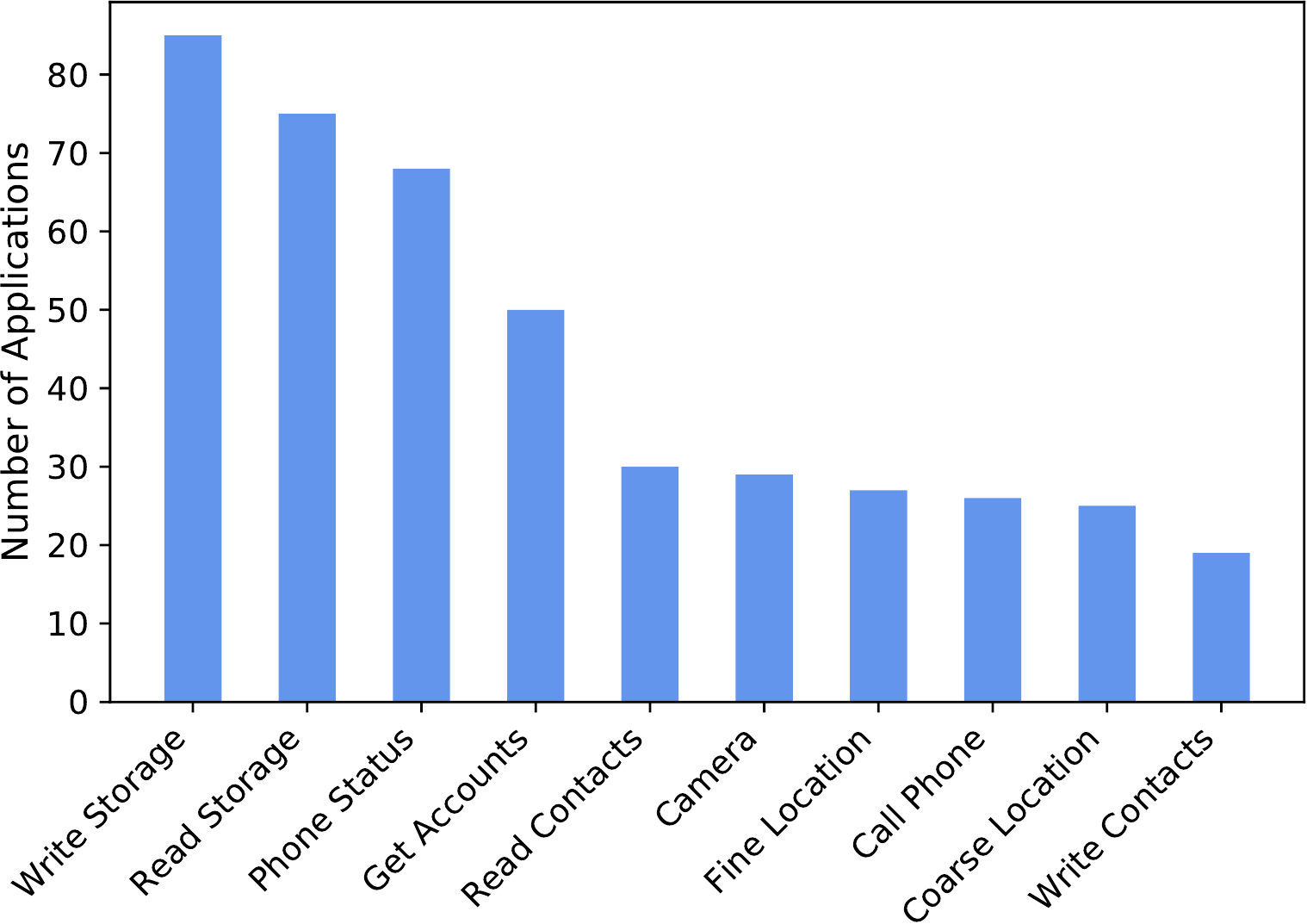}
\caption{Top 10 Dangerous Permissions}
\label{fig:Dang_top_10_permissions}
\end{figure}

We found that 35 different dangerous permissions were requested by the selected 100 security apps. Top-10 frequently requested permissions are shown in Figure~\ref{fig:Dang_top_10_permissions}. As can be seen from the figure, the two most frequently requested permissions were \texttt{WRITE\_EXTERNAL\_STORAGE} and \texttt{READ\_EXTERNAL\_ STORAGE}. The purpose of these two permissions is to read and write from external storage such as SD cards. More often than not, mobile users store personal information and data such as photos, documents, videos on external storage. Thus, despite the legitimate uses of these permissions in the likes of scanning for viruses or deleting malware and unused files, a bad actor can abuse this permission to steal personal information.

% , if the user allowed those two permissions, the applications can get access to the content stored in the phone and potentially share with third parties.

%\textcolor{blue}{The purpose of these two permissions is to read and write from external storage such as SD cards. More often than not, a mobile user stores personal information and data such as photos, documents, videos on external storage to avoid memory limitation issue in a device. Therefore, these two permissions are considered dangerous and must not be accessed unless the purpose of using these permissions are explicitly stated in the policy. In addition, apps may also use external storage locations to store sensitive information. For example, an app can store a configuration file in public storage that can be modified by a third- party app on the device. If the file contains information about user authentication and other resources used to process the app, then a malicious app on a device can harvest such information and use it for illegitimate purposes such as phishing.} 
 
%  \textcolor{red}{TODO: Can we explain why? I am guessing reading is for virus scanning but not sure about writing. Usually security app websites describe why they need this permission in their pages.}

Similarly, another frequently asked permission \texttt{READ\_PHONE\_STATE} enables security features such as call blocking or blacklisting. However, granting this permission will also allow access to read the phone number or interrupt legitimate calls. In the privacy policy analysis we found that some applications collect the users’ phone number and they may use it for advertising and even share with third parties. Such information at wrong hands can cause major damages to the user such as identity theft.

% Moreover, this permission not only allow accessing the phone number but also the network information and call status. \textcolor{red}{TODO: I don't get what is meant by call status} 
 %Thirty different dangerous permissions are requested by 100 Anti-Virus applications with the frequency in the range of 1 to 85. Similar to the distribution of normal permission, ``WRITE\_EXTERNAL\_STORAGE'' permission and ``READ\_EXTERNAL\_ STORAGE'' permission is requested by 85 times and 73 times separately. 
 
 %The ``GET\_ACCOUNTS'' permission is requested by 50 times; half of the applications have used this permission to get the list of accounts. Figure~\ref{fig:Dang_top_10_permissions} shows the distribution of dangerous permissions.

% without risks. Therefore, more attention should be put on the analysis of dangerous permissions. The data requested by dangerous permission has threats to users’ personal information, which have threats in storage data and other applications.
%Theoretically, all normal permission can only get access to the data in the outside of sandbox 

To summarise, the majority of the requested dangerous permissions have an associated legitimate security feature included in the security app. However, in our privacy policy analysis we found evidence the collected data may have other uses and can be shared with third parties. As such, the security app users need to be cautious when granting such permissions and need to carefully evaluate the trade-offs between the feature and the potential privacy threats.

\section{Malware Detection}
\label{Sec:Malware}

\begin{table*}[!t]
\scriptsize
\caption{The description of Six Selected Malware/Virus}
\label{tab:my-table}
\centering
\begin{tabular}{p{1.5cm}|p{1.5cm}|p{2.5cm}|p{9cm}}

\hline
\textbf{Malware Name} & \textbf{Release Year} & \textbf{Makeup} & \textbf{Malicious Behavior} \\ \hline
Trojan & 2017 & Fake Adobe Flash Player update the web page & 1. Once installed, it pops up  a dialogue box requesting a device information from a user.\\
& & & 2. If a user grants access to device information then this malware contacts command and control (C\&C) server and sends device information for possible exploitation. \\ \hline
Backdoor & 2015 & BeNews app (Already removed from the Google Play Store) & 1. This malware successfully bypasses the security verification check from the Google Play Store and has been downloaded over fifty times by the device users before it is removed from the store. \\
& & & 2. The malware works by opening the backdoor for hackers to remotely control the victim's device and installing different types of malware(s) on a device, once a user clicks the app. \\ \hline
Spyware & 2016 & Banker Trojan (app name is various time to time) & 1. Once installed on a mobile device, this malware hides its icon (in a stealth mode) and keeps sending the device information to the C\&C server. \\
& & &  2. This malware type also tricks user to enter or update their credit card details into a fake Google Play web page. \\ \hline
Banking malware & 2016 & Sberbank (a reputable Russian bank online app) & 1. This malware attempts to perform a phishing attack on a user by creating an app logo that looks similar to the banking apps icon. Hence, it misleads a user to download the app and steals his credentials. \\
&  &  & 2. The malware is also capable of intercepting SMS messages and incoming calls on a mobile device. \\ \hline
Banking malware (new) & 2019 & BatterySaverMobi & 1. This malware was downloaded over one thousands times with dozens of fake five-star ratings at the Google Play Store. \\
&  &  & 2. The malware gets a device administrator privileges by luring users to install a fake system update. \\
&  &  & 3. It collects users banking information through inbuilt keylogger module and screenshots, and is much flexible in hiding itself than other similar Trojans. \\ \hline
Xbot & 2016 & Fake Google Plays payment interface & 1. This malware imitates Google Play Store payment interface alongwith the login pages of seven banks to steal banking credentials and credit card information. \\
&  &  & 2. The malware is also capable of control infected devices remotely.  \\
&  &  & 3. It also encrypts victim's files and ask for the ransom to decrypt them. \\
\hline
\end{tabular}
\end{table*}

%\subsection{Performance Evaluation}\

We next evaluate how effective are these security apps in detecting known Android malware. We selected six malware samples that have been detected and publicly disclosed between 2015 and 2019. All sample malware are sourced from GitHub, and the general description of each selected malware is shown in Table~\ref{tab:my-table}. 

Using these malware we tested two scenarios; \textit{i) whether the security apps can detect a copy of malware stored in a  phone} and \textit{ii) whether the security apps can detect malware installed in the phone}. We installed the selected security apps one by one in a Google Pixel phone and assessed its performance with respect to these two tasks. We highlight that we could test only 86 out of the 100 apps we selected because 14 of them were limited to Samsung phones only.

\begin{figure}[!htbp]
\captionsetup{skip=0pt, justification=centering}
\centering
      \includegraphics[scale= 0.45, keepaspectratio]{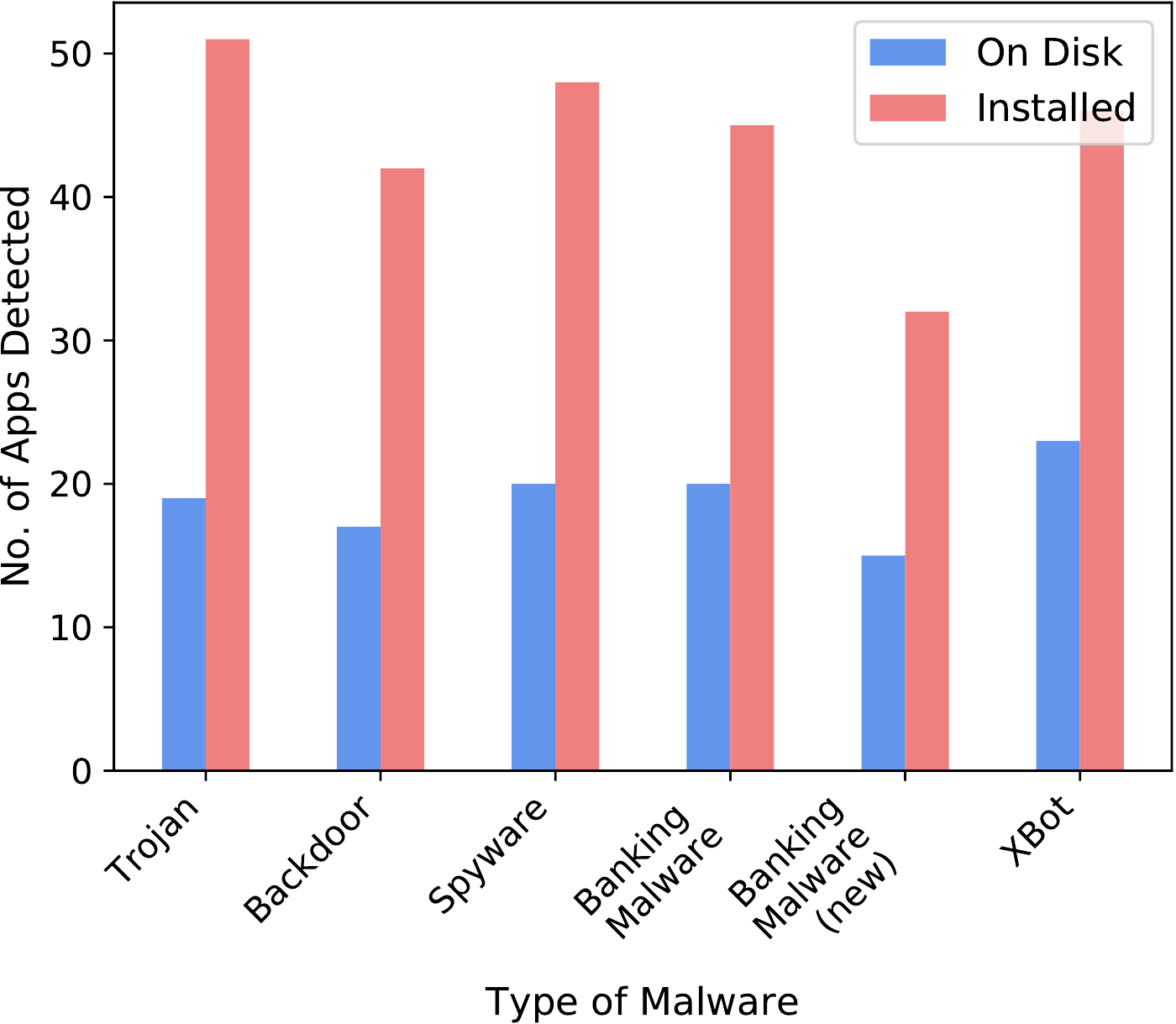} 
\caption{Detection rates of various malware types}
\label{fig:12}
\end{figure}

We show the results in Figure~\ref{fig:12}. Surprisingly, detection rate of disk copies were quite low. For each malware sample, only 15-20 of the security apps could detect them even after running a full system scan (when such a feature is available). Malicious file detection can be considered as the first line of defence when it comes to defending smartphones and lack of that indicates a serious functional limitation in commercial security apps.  

The detection rates installed malware was high compared to disk copies, yet not perfect. Between 40-50 security apps out of 86, were able to identify installed malware except for the case of the  banking malware. The detection rate of the banking malware for both on disk file and installed file was much lower than the others. The main reason for this may be the fact that this malware was released in 2019 whereas the others are a few years older. \textit{Nonetheless, it is an eye-opening result that 20-30 so called security apps fail to identify actual malware installed in a smartphone and also show much lower detection rates when it comes to more recent, yet publicly known malware. This gives a false sense of security to the users  who are under the impression that their devices are safe due to the presence of a security app.}

Further to this, as shown in Table~\ref{tab:3}, 30 apps were not able to detect any of the installed malware and 12 detected only 1-3 malware samples. Only 32 were able to detect all six samples of installed malware. Example apps that did not identify app six malware samples includes ``SandBlast Mobile Protect'' and  ``Free Spyware Removal Info'',  while apps such as ``Norton'', ``McAfee'', ``Kaspersky'', and ``Safe Security'' were able to identify all six samples of the malware. \textit{We highlight that one factor behind lower detection rates is that there are some apps that provide different security features than virus and spyware scanning.}

% \textcolor{blue}{example of cannot detect virus: com.lacoon.security.fox; com.example.antivirus.software; com.appsplan1.freespywareremovalinfo.apk; com.ashampoo.rottensyschecker.apk}. 

% \textcolor{blue}{example of can detect all virus: antivirus.maka.scannner.security.apk; antivirus.virusremoval.security.phone.apk; com.msysoft.viruscleaner.security; com.qihoo.security}. 
%.The result also shows that the installed malware detection rate is much higher and the detection rate is achieved averagely 50\% (43 of 86) of each group except one group named banking malware (new). 

\section{Network Traffic}
\label{Sec:Tracking}

Since the majority of the security app we investigated (95 out of 100), it is possible these apps might monetise user data through advertising and analytics. We already reported such evidence during our privacy policy analysis (cf. Section~\ref{Sec:PrivacyPolicy}). To investigate further on this, we used the \textit{Lumen privacy monitor}~\cite{razaghpanah2018apps} which  can continuously monitor all installed applications' operations and record all traffic flows of each application separately. Lumen also can identify when personal information is leaked by a particular application. 

To monitor all the traffic flows from each application, we installed applications one by one and set the monitor time to six hours. We granted all the permissions requested by the app so that it is fully functional. Next we extracted Lumen's dataset for further analysis.

\begin{table}[!t]
\centering
\caption{Detection of installed malware}
\label{tab:3}
\begin{tabular}{c|c} \hline

{\bf No. of detected malware} & {\bf No. of detecting security apps} \\ \hline
 
0 & 30  \\ 
1 - 3 & 12   \\ 
4 - 5 & 11 \\ 
6 & 32  \\ \hline
\end{tabular}
\end{table}

\subsection{Packet Destinations: IPs, Ports, and Domains}

% \begin{figure}[!t]
% \captionsetup{skip=0pt, justification=centering}
% \centering
%       \includegraphics[width=\columnwidth, keepaspectratio]{figures/Application_Ports.pdf} 
% \caption{Distribution of Destination Ports}
% \label{fig:15}
% \end{figure}

\begin{figure}[!h]
\centering
%\label{fig:phished_web}
      \includegraphics[scale=0.45]{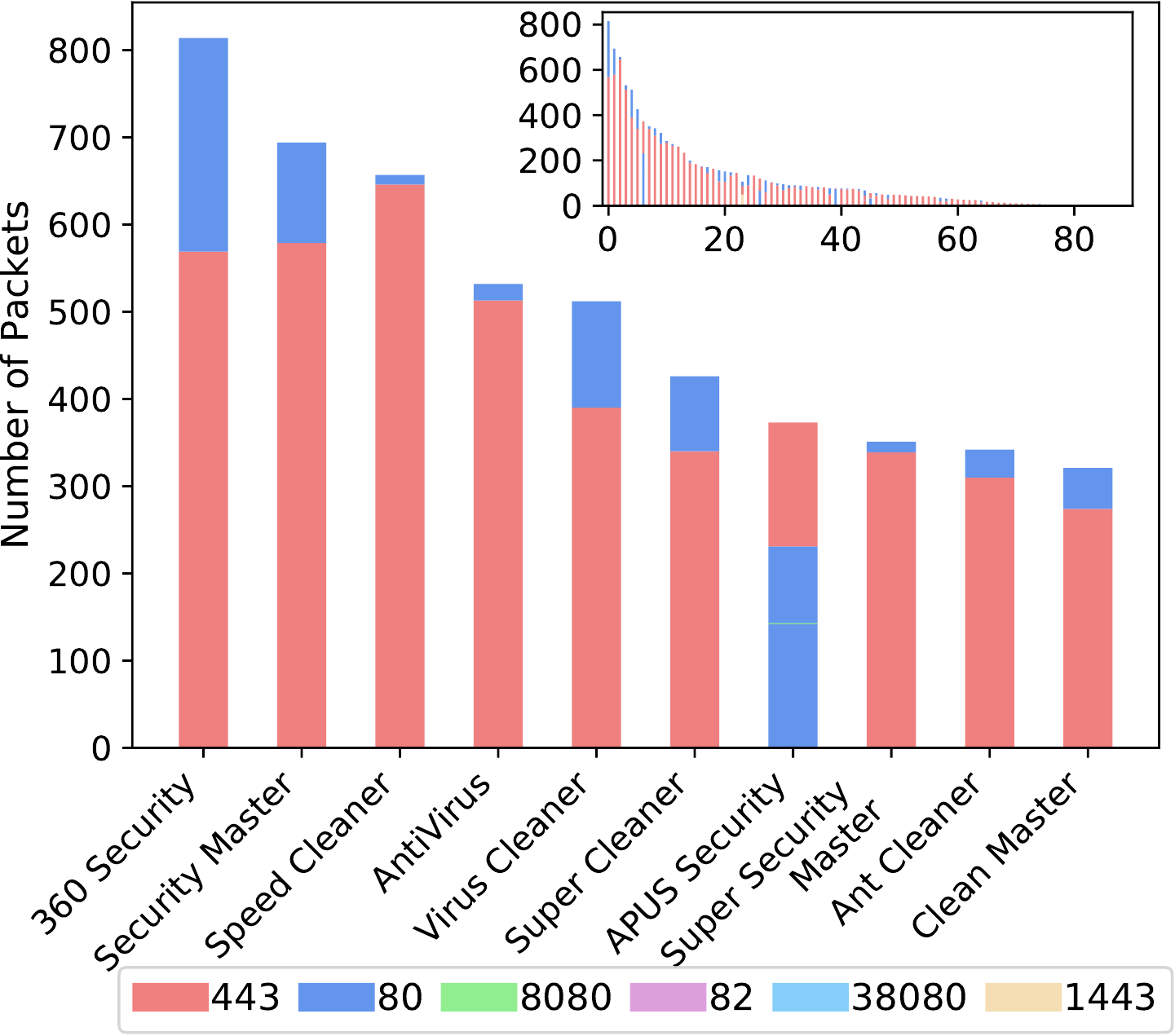}
\caption{Distribution of Packets Sent by Apps}
\label{fig:16}
\end{figure}

In Figure~\ref{fig:16} we show the number of packets send by each app during the six hour observation period to different destination ports (top-10 according to number of packets as well as all apps). 360 Security is the app that sent out the highest number of packets. Majority of the apps sent encrypted packets to port 443 with the exception of APUS Security, which had a significant fraction of non-encrypted traffic going into port 80. Other apps with noticeable unencrypted traffic included 360 Security, Security Master, Virus Cleaner, and Super Cleaner.

% \begin{figure}[!t]
% \captionsetup{skip=0pt, justification=centering}
% \centering
% \subfloat[Top 10 Encrypted and Un-encrypted Packets Distributions by Apps]{\label{fig:enctop10PacketDist}
%       \includegraphics[width=1.0\columnwidth, keepaspectratio]{figures/Top10_Encrypted_pckt.pdf}} \\
% \subfloat[Encrypted and Un-encrypted Packets Distribution of All Apps] {\label{fig:encpcktDistribution}
%       \includegraphics[width=1.0\columnwidth, keepaspectratio]{figures/Bar_Encrypted_Packets.pdf}}
% \caption{Distribution of Packets w.r.t Encrypted Traffic Flow}
% \label{fig:17}
% \end{figure}

\begin{figure}[!h]
\centering
%\label{fig:phished_web}
      \includegraphics[scale=0.45]{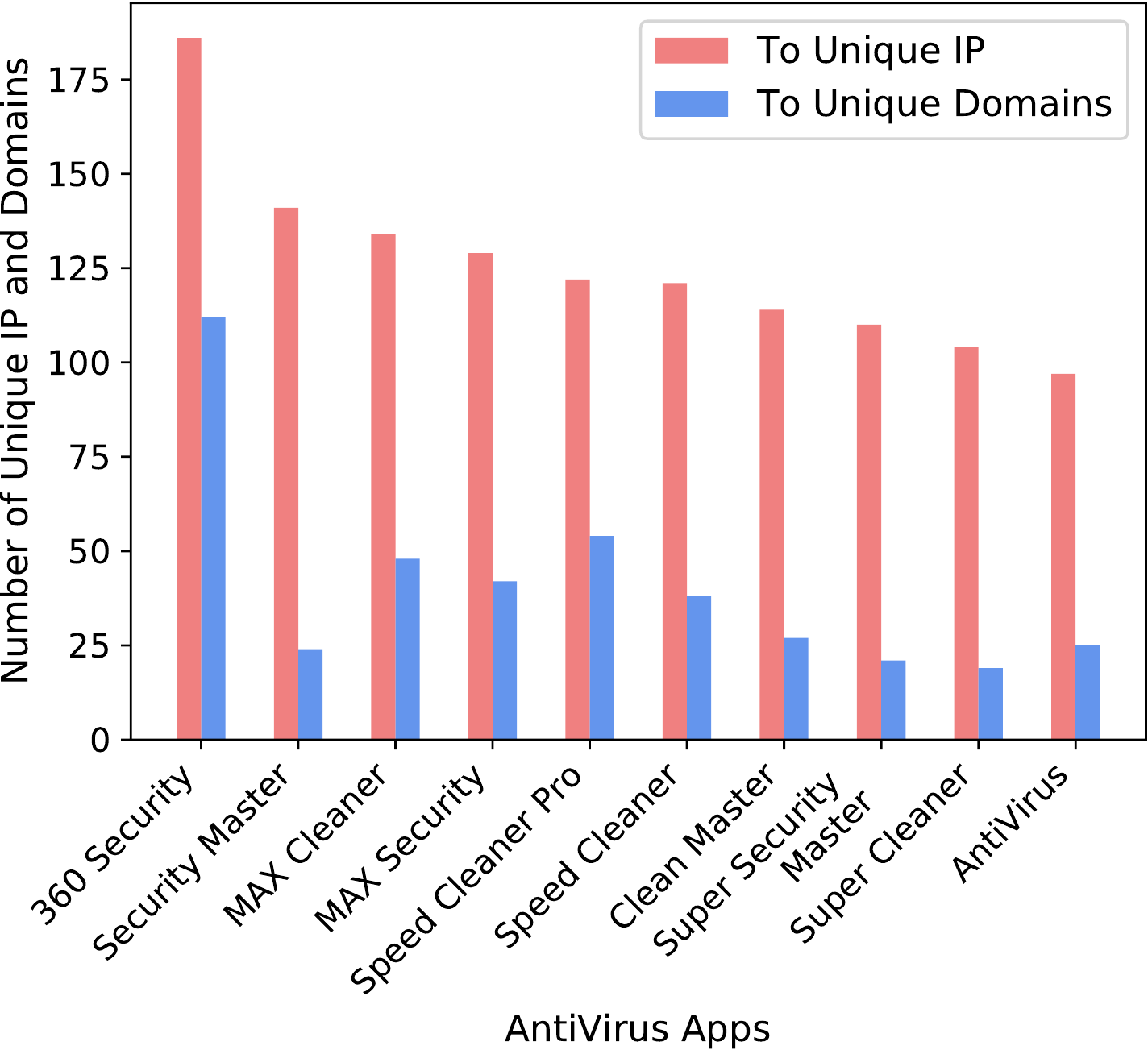}
\caption{Distribution of Unique IP Addresses and Domains}
\label{fig:18}
\end{figure}

Next, in Figure~\ref{fig:18} we show the top-10 apps according to the number of unique IP addresses they contact. 360 Security was again at the top with contacting over 180 unique IP addresses accounting for over 100 domains. In summary, there were nine apps that contacted over 100 IPs and three apps that contacted over 50 domain names.
% \textcolor{red}{TODO: Do we have the correct numbers for this?}

%The data extracted from the Lumen also can track the destination IP and Domain name that each application sends packages to. Figure~\ref{} indicates how many unique IP and Domain names are identified as traffic package destination from each application, and only the top 10 application will be listed in the graph. Each of the top 10 applications has sent packages to over 100 IP addresses and multiple domains. 360 Security application is ranked as the top one application that distributes its packages to over 180 IP addresses (over 100 Domains). Consider both the number of packages and the number of unique IP sent by 360 Security are extremely large, 360 Security will be picked for further analysis.

% \begin{figure}[h!]
% \captionsetup{skip=0pt, justification=centering}
% \centering
%       \includegraphics[width=\columnwidth, keepaspectratio]{figures/Top10_IP_Domain.pdf} 
% \caption{Distribution of Unique IP Addresses and Domains \textcolor{red}{TODO: Break x-ticks more naturally, unprintable app name /change the colors to usual scheme}}
% \label{fig:18}
% \end{figure}

\begin{figure}[h]
\centering
%\label{fig:phished_web}
      \includegraphics[scale=0.45]{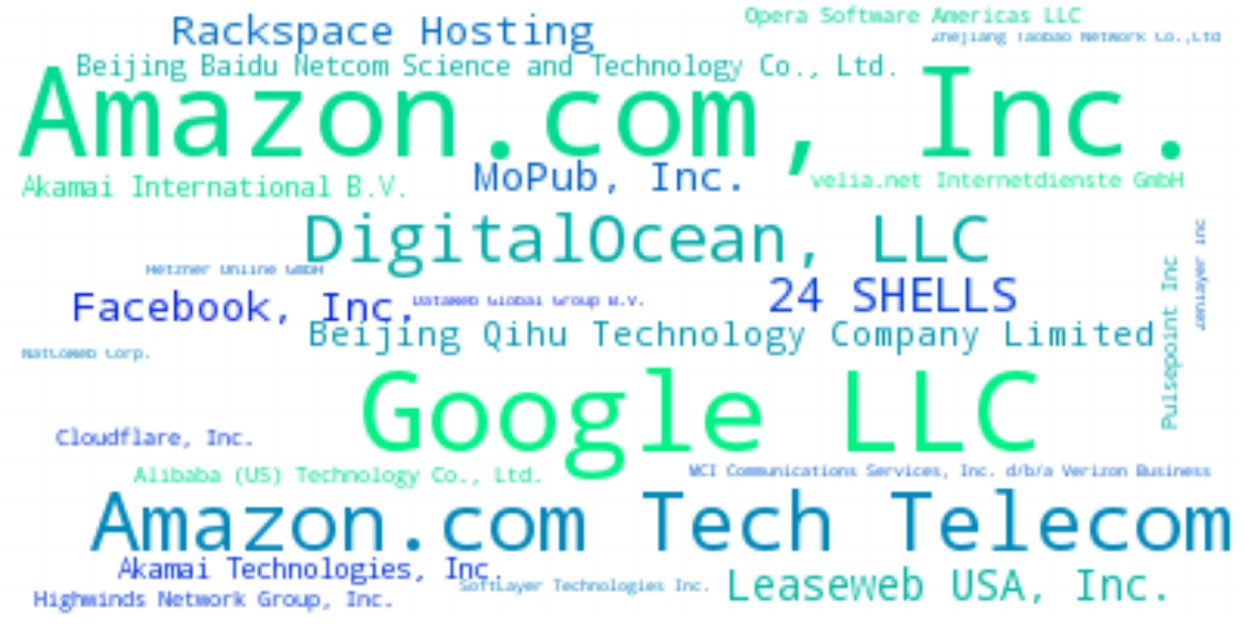}
\caption{Top domain names contacted by the app 360 Security}
\label{fig:domaincloud}
\end{figure}

Most frequent domains that are being contacted by 360 Security are visualised in Figure~\ref{fig:domaincloud}. While the majority of the traffic has gone to cloud service providers such as Amazon, Google, DigitalOcean, and Rackspace, we also see the advertising company Mopub's domain name among the domains indicating possible advertising or analytics activities. 

%It is obvious that 360 Security has sent most of the packages to Amazon and Google. 360 Security may accept the web service provided by Amazon or Google; this causes the frequent communication between them. The geographic information of several outstanding domain names are spotted on the map (Figure~\ref{fig:geomap}). The most domains are located on the United States, and some packages are sent to a domain named Beijing Baidu that located in China.    

% \textcolor{red}{Can we redraw this to enhance the color of the background text. That cream color is barely visible. Can we generate a similar figure for all the domains of all the apps?}

% \begin{figure}[!t]
% \captionsetup{skip=0pt, justification=centering}
% \centering
% \subfloat[Cloud of Domain Names]{\label{fig:domaincloud}
%       \includegraphics[width=1.0\columnwidth, keepaspectratio]{figures/360IP.png}} \\
% % \subfloat[Domain Name's Geographic location] {\label{fig:geomap}
% %       \includegraphics[width=1.0\columnwidth, keepaspectratio]{figures/360IPMap.png}}
% \caption{Distribution of Geographical of Apps and Linked Packets}
% \label{fig:19}
% \end{figure}

\begin{figure}[!h]
\centering
%\label{fig:phished_web}
      \includegraphics[scale=0.45]{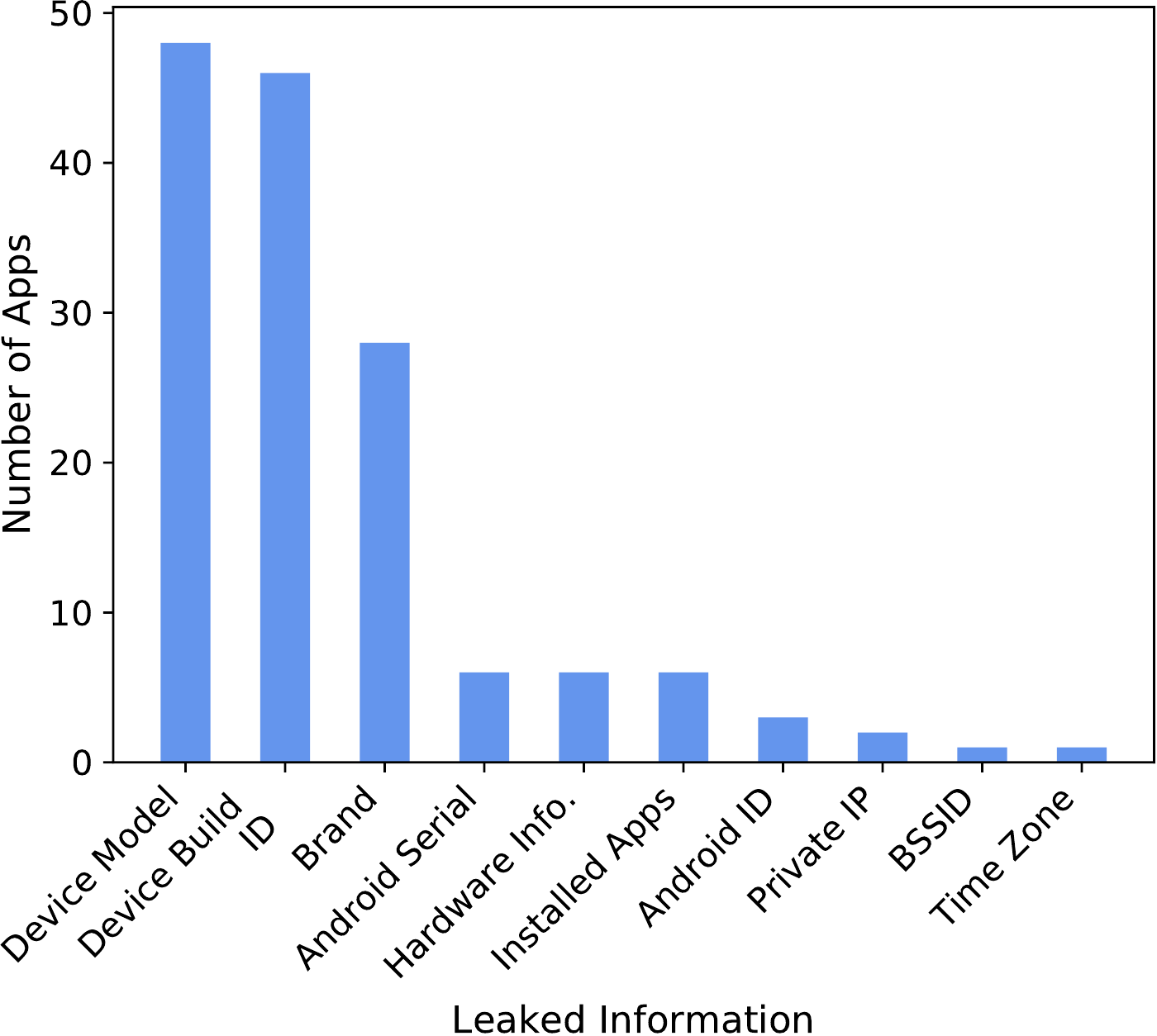}
\caption{Personal information extracted by security apps}
\label{fig:20}
\end{figure}

\subsection{Personal Information Leakage}

Finally, we investigate what type of personal information are being collected by the security apps according to the Lumen report. Figure~\ref{fig:20} summarises the findings. The top three frequently collected information are device model, build id, and device brand. Such types of information are potentially collected for record keeping (e.g. a virus was found only in certain brands of phones). Nonetheless, we also found a limited number of apps that collect more personal information in nature such as list of installed apps and personally identifiable information such as the Android serial, Android ID, and BSSID. While such information collection can still have legitimate uses in the context of device security, leaving such information in the hands of third-parties can pose significant privacy threats to the users~\cite{seneviratne2014predicting}\cite{seneviratne2015your}. 

\textit{All in all, it is surprising to see multiple security apps still not using fully encrypted connections. Also, it is questionable whether all of these network connections are essential for the core features of the app. Finally, we found evidence of security apps indeed collecting and transmitting personal information to their back-end servers and users need to be wary of this, given our findings on privacy policies.}

%After summarizing the privacy policy declared from each application, it can consider that the possible reason for collecting end users’ information is used to keep updating their user group statistics and improving their services. However, the graph only displays all types of collected end users’ information without explaining how the collected data is used. It needs further analysis in the future to verify if the real process of collected information is same as declared purposes.

% \begin{figure}[!t]
% \captionsetup{skip=0pt, justification=centering}
% \centering
%       \includegraphics[width=\columnwidth, keepaspectratio]{figures/Leaked_Info.pdf} 
% \caption{Sensitive Information Extracted by AntiVirus Apps \textcolor{red}{TODO natural x-tick breaks, Build ID, Time Zone, Serial No, pii\_type not clear}}
% \label{fig:20}
% \end{figure}

\section{Discussion and Concluding Remarks}
\label{Sec:Discussion}

In this paper we presented the first multi-faceted study of Android security apps. We next discuss the implications of our findings, limitations, and possible extensions. \\ \vspace{-2mm}

\noindent{{\bf Providing a false sense of security}} - One of our key findings was that a significant fraction of Android security apps provide a false sense of security. Around 75\% of the apps failed to identify malware copied into the phone storage. While the detection rates were better when it comes to installed malware ($\sim$50\%), they were not ideal. Also, we noted that detection rates drops further when it comes to very recent malware. Therefore, the users must not by any means consider their devices secure due to the mere fact that they have installed an Android security app. They must read the product reviews and understand the features provided by these apps. Most importantly, users must understand apps' limitations as well as the  features not provided by these apps. \\ \vspace{-2mm}

\noindent{{\bf Not following best software development practices}} - We also found evidence of developers not following best software development practices. For example, multiple apps were generating significant volumes of unencrypted traffic and there were a few apps without privacy policies. Also, as highlighted earlier some virus databases are not updated frequently. While both Google and Apple takes active measures~\cite{mobilemarketer}\cite{iosvirus} to control fraudulent security apps, it appears that much stricter checks are required.    \\ \vspace{-2mm}

\noindent{{\bf Personal information collection and usage}} - Through permission analysis we demonstrated that the security apps have access to vast amount of personal data. Our privacy policy analysis showed that the majority of the security apps do indeed collect personal information and we found empirical evidence of such data collections in network traces as well. The users must be wary of these data collections and must carefully assess the trade-offs between the security features they require and the possible personal data leakage. It is important to read the fine-print of the privacy policies and understand how the collected data is used. Finally, to assist the end users in making their decisions, we believe it is a good practice for the app developers to publish descriptions of the uses of various permission requests.  \\ \vspace{-2mm}

\noindent{{\bf Limitations and future work}} - Our analysis can be further extended in multiple ways. For example, further analysis can be conducted to trace what data is going to what domains by using an SSL proxy. That will allow to further segregate and decide whether the data is collected for the core features of the app or for advertising purposes. Also, privacy policies can be studied together with actual data collections, along side with privacy law experts and investigate whether the apps comply the applicable privacy laws and regulations. Permission analysis can be further expanded by decompiling the codes and analyse the actual API calls that are being called. This will allow to obtain a much fine-granular understanding of apps behaviours. In this work, we investigated only the effectiveness of the security apps in detecting malware. However, among the apps we investigated, there were apps that did not claim to provide malware scanning features, but other security features. Thus, further analysis can be done to investigate the performance of the security apps with respect to their specific features.   \\ \vspace{-2mm}

%\input{conclusion}

% can use a bibliography generated by BibTeX as a .bbl file
% BibTeX documentation can be easily obtained at:
% http://www.ctan.org/tex-archive/biblio/bibtex/contrib/doc/
% The IEEEtran BibTeX style support page is at:
% http://www.michaelshell.org/tex/ieeetran/bibtex/
\bibliographystyle{IEEEtranS}
% argument is your BibTeX string definitions and bibliography database(s)
\bibliography{references}

\end{document}